%% file: paper.tex
\documentclass[manuscript,screen,final]{acmart}
\AtBeginDocument{%
  }

\setcopyright{acmlicensed}
\copyrightyear{2026}
\acmYear{2026}
\acmDOI{XXXXXXX.XXXXXXX}

\begin{document}
\title{Designing Medical Chatbots where Accuracy and Acceptability are in Conflict: An Exploratory, Vignette-based Study in Urban India}

\author{Ananditha Raghunath}
\affiliation{%
  \institution{University of Washington}
  \city{Seattle}
  \state{Washington}
  \country{USA}}
\email{araghu@cs.washington.edu}

\author{William Thies}
\affiliation{%
  \institution{Everwell Health Solutions}
  \city{Bangalore}
  \country{India}}
\email{bill@everwell.org}

\author{Mohit Jain}
\affiliation{%
  \institution{Microsoft Research India}
  \city{Bangalore}
  \country{India}}
\email{mohja@microsoft.com}

\renewcommand{\shortauthors}{Raghunath et al.}

\begin{abstract}
When medical chatbots provide advice that conflicts with users’ lived care experiences, users are left to interpret, negotiate, and evaluate the legitimacy of that guidance. In India, the widespread overuse of antibiotics, antidiarrheals, and injections has shifted patient expectations away from the guideline-aligned advice that chatbots are trained to provide. We present a mixed-methods, vignette-based study with 200 urban Indian adults examining preferences for and against guideline-aligned, norm-divergent advice in chatbot transcripts. We find that a majority of users reject such advice, drawing on diverse rationales grounded in their lived expectations. Through the design and introduction of context-aware nudges, we support expectation alignment that shifts preferences towards transcripts containing guideline-aligned advice. In doing so, we surface key tensions in the equitable design of medical chatbots in the Global South.
\end{abstract}

\ccsdesc[500]{Human-centered computing~Empirical studies in HCI}
\ccsdesc[500]{Applied computing~Health care information systems}
\ccsdesc[300]{Social and professional topics~Geographic characteristics}

\keywords{medical chatbots, clinical guidelines, context-aware nudges, healthcare access, Global South}
\maketitle

\input{sections/1-intro}

\input{sections/2-rw}
\input{sections/3-methods}

\input{sections/4-results}
\input{sections/5-discussion}

\input{sections/6-conclusion}

\bibliographystyle{ACM-Reference-Format}
\bibliography{ref}

\end{document}

%% file: sections/1-intro.tex
\section{Introduction}
Medical chatbots are increasingly considered a viable means of providing timely, quality care to patients in high-volume settings (e.g. \cite{inkster2018wysarealworld, ramjee2024cataractbot}). The prevailing method of equipping bots with medical knowledge is by curating knowledge corpora composed of formal medical guidelines, which large language models (LLMs) can utilize for fine-tuning or data retrieval to generate clinically-valid responses to patient questions (e.g. \cite{singhal2022-medpalm}). Strict adherence to guidelines is emphasized during development and while obtaining regulatory clearance, and many bots demonstrate such adherence through high scores on medical licensing exams, such as the United States Medical Licensing Exam (USMLE) \cite{singhal2022-medpalm, gpt4-usmle}. 

The latent assumption underlying this approach is that once chatbots reach sufficient accuracy, patients can safely access quality medical advice and, in turn, improve their health outcomes. However, sociotechnical interactions are never that straightforward. In many regions, such as India, common medical treatments have been widely documented to deviate from clinical guidelines, creating a potential dissonance between patients’ lived care experiences and chatbot-provided advice. A 2024 study in urban India found that up to 45\% of prescriptions across 13 tertiary care centers did not comply with standard treatment recommendations \cite{Shetty2024Evaluation}. To set the sociotechnical context for our work, we describe guideline-divergent treatment norms in India related to antibiotics, antimotility drugs, and injections in the paragraphs below.

Antibiotics have been documented to be widely overprescribed in India for conditions where they offer no therapeutic benefit. A 2020 National Center for Disease Control study found that antibiotics were given prophylactically in 55\% of patient interactions across 20 Indian tertiary care centers, while another study found that up to 88\% of surveyed  doctors self-reported prescribing antibiotics to patients with viral infections where they provide no therapeutic benefit \cite{NCDC2024NACNET, Nair2019Without}. Patients are known to subsequently reuse the incorrect prescriptions at pharmacies to obtain additional antibiotics when experiencing similar symptoms \cite{Kotwani2021ConsumersAntibiotics}. Such antibiotic overuse contributes to antimicrobial resistance, which directly causes over 1 million deaths and contributes to approximately 5 million deaths annually worldwide \cite{WHO_AMR_2023}. 

Antidiarrheal medications have also been documented to be overused in Indian children. A 2016 study found that 18–62\% of children receive inappropriate antimotility drugs or antibiotics when they present with diarrheal symptoms. This practice increases costs, adverse reactions, and even childhood deaths \cite{Walker2016ManagementDiarrhea}. 

The use of injections has also been shown to diverge substantially from clinical guidelines, with a study finding that 61\% of outpatients received injections despite clinical indications in only 5–10\% of cases. Unsafe injection practices and injection reuse expose patients to the risk of contracting Hepatitis B, HIV/AIDS and other diseases \cite{Janjua2016SafeInjection}. Additionally, the overuse is disproportionately experienced by rural patients (73\%) and those from tribal communities (86\%), highlighting a significant equity concern associated with this norm \cite{TNN2009Injections}. 

Taken together, this body of evidence suggests that patient treatment expectations in the Indian sociocultural context may not align with the guideline-aligned advice provided by medical chatbots. To examine how users make sense of the tension between guidelines-based advice and diverse care experiences, we leverage a two-phase, vignette-based study. We ask the following \textbf{RQ1}: In settings where lived care experiences and guideline-aligned care may diverge, do patients prefer a chatbot that provides norm-congruent but guideline-divergent advice? With the support of expert clinicians, we construct vignettes for three common conditions where lived care experiences may diverge from guideline-aligned care: the common cold (antibiotics), viral diarrhea (antimotility drugs), and tension headache (injections). In Phase 1, we simulate two medical chatbots, creating transcripts that arrive at the same, correct diagnosis based on the symptoms from the vignette. Following diagnosis, the first chatbot, \textit{Verity}, concludes the interaction with a guideline-aligned treatment recommendation. The second chatbot, \textit{Max}, suggests norm-congruent but guideline-divergent treatment. We present the chat transcripts to 100 participants, finding that a majority of participants prefer \textit{Max} over \textit{Verity}. We complement this result with qualitative analysis to surface how participants articulate expectations and legitimacy when evaluating advice. 

Given this understanding, in Phase 2, we investigate the impact of a design mechanism aimed at supporting expectation alignment, asking: \textbf{RQ2}: Can embedding a \textit{context-aware nudge} into a chatbot's dialogue, i.e., wrapping the guideline-aligned advice with extra information that is aware of the context in which it's being interpreted, support user interpretation? We show 100 new participants the transcript from Max in Phase 1, and a transcript from a new bot named \textit{Clarity}, which incorporates a context-aware nudge preceding guideline-aligned advice. We analyze both quantitative and qualitative responses in this phase, finding that nudges shifts participants’ sensemaking processes and, in turn, reorient preferences toward \textit{Clarity} across educational groups.

Together, our work examines how users accept, evaluate, and negotiate medical advice from chatbots in contexts where clinical guidelines diverge from lived treatment norms. In the sections that follow, we describe our study design and methods, present findings from each phase, and discuss implications for the design of medical AI systems that seek to balance clinical accuracy with user acceptability in Global South contexts.

%% file: sections/2-rw.tex
\section{Related Work}
Numerous Human Computer Interaction (HCI) studies have examined the design and use of AI-based health chatbots for underserved populations. In Global South contexts, prior work has explored chatbots for addressing COVID-19 vaccine hesitancy ~\cite{vaccine-hesitancy-cscw24}, supporting breastfeeding education~\cite{yadav-feeding-cscw19}, promoting adolescent sexual and reproductive health~\cite{Deva2025Integrating, AdolescentBot-chi21, SnehAI-JMIR22, nthabi-resSq23}, facilitating information gathering prior to clinical consultations ~\cite{waitingroom-chi24}, mediating self-disclosure in mental health care ~\cite{mediator-mental-cscw20}, among others ~\cite{elder-bot-cscw20, issue-bot-health-tiis22}.

Prior work shows that users do not passively accept advice from chatbots, but actively evaluate their credibility, authority, and appropriateness in context ~\cite{moral-agency-cscw24, human-side-bot-21, seitz22can}. Their judgments are by technical characteristics such as information quality and explainability, but also by perceived risk and relational qualities such as anthropomorphism and working alliance ~\cite{wutz23factors}. Studies situated in India and other Global South contexts further emphasize that trust in medical AI is embedded within broader sociotechnical care ecosystems. Users’ expectations and evaluations are influenced by service quality, surrounding care practices, and culturally situated expectations of what constitutes “good care” ~\cite{Wadhwa2025DesigningWithCulture, prakash24why, nimisha-bp-tochi}. 

Related work demonstrates that adapting health technologies to users’ sociocultural contexts—through tailored communication, respectful engagement with local beliefs, or health-literacy-sensitive framing can improve legitimacy and engagement ~\cite{bot-literacy-10, AdolescentBot-chi21, taboo-chi17, sugar-weekend-chi19}. While some work on algorithmic authority shows that users actively negotiate the epistemic status of algorithmic systems, less attention has been paid to what happens when users’ lived care norms diverge from recommended clinical care, or to cases where norm-conforming but clinically inaccurate advice may be perceived as more legitimate. We extend this body of work by conceptualizing legitimacy as an design achievement, negotiated moment-by-moment through how systems acknowledge, align with, or challenge users’ lived care norms.

Nudges were first articulated within behavioral economics as subtle, non-coercive interventions that shape decision-making~\cite{thaler21nudge}. They have been widely studied across domains, including digital systems and HCI ~\cite{beshears20nudging}. Prior HCI work has catalogued a broad range of nudging mechanisms, showing how interface-level cues leverage cognitive tendencies to influence interpretation and choice ~\cite{caraban1923, fogg-persuasion, hamari14persuasive}. In health contexts, design has frequently been used to encourage adherence and self-management, often by establishing system credibility through expert endorsement, citations, or institutional affiliation~\cite{hamari14persuasive}. While such approaches have demonstrated benefits for influencing behavior, prior work exhibits two key limitations. First, reviews of digital nudging highlight that much of this literature relies on controlled or online experiments, with comparatively fewer studies examining how nudges function in situated, real-world contexts~\cite{bergram22digital}. Second, nudges are largely framed as mechanisms for promoting compliance with recommended actions (e.g., increasing physical activity). In contrast, our work introduces \textit{context-aware nudges} not as instruments of compliance, but as resources that support expectation alignment between users’ local care contexts and a chatbot’s guideline-aligned dialogue. We study their impact through an in-person study in urban India.

We conduct our in-person study using vignettes. Vignettes are “sketches of fictional scenarios” presented to respondents, who are “invited to imagine, drawing on their own experience, how the central character in the scenario will behave.” As such, vignettes enable the collection of situated data on group values, beliefs, and behavioral norms~\cite{BloorWood2006Keywords}. Vignette-based studies are a well-established method in HCI and health research for examining how users react to, interpret, reason about, and respond to complex or sensitive scenarios (e.g.,~\cite{vignette1, Aoki2025VignettesLMIC}). We extend this body of work by embedding hypothetical health-seeking and care provision scenarios within chatbot transcripts, retaining experimental control while  measuring preferences, expectations and sensemaking processes. 

%% file: sections/3-methods.tex
\section{Methods}
Below, we document the methods used in our two phase, mixed method study. 

\subsection{Study Context and Participants} 
We conducted the study in July 2024 at a large institutional campus in urban Bangalore, India. The campus employs over 1000+ staff members across a wide range of roles, including as researchers, interns, engineers, administrators, security personnel, custodial staff, kitchen staff, drivers, etc. This setting offered an opportunity to study responses to medical advice among participants with shared geographic and institutional context, but divergent educational backgrounds, socioeconomic positions, and experiences with healthcare. We recruited two cohorts of 100 participants, one for each study phase, using face-to-face convenience sampling across eight days. To reduce temporal and spatial sampling bias, recruitment occurred at different times of day and across multiple locations on campus across eight days. Inclusion criteria were being over 18, affiliated with the institution, and ability to communicate in English, Hindi, Tamil, or Kannada. We used educational attainment as a structural proxy for differential access to formal education and health-related information, distinguishing between participants with educational attainment up through grade 12 (Lower Education subgroup) and those with bachelor’s degree or above (Higher Education subgroup). We documented parental status as it could provide context for a participant’s response in one of our clinical vignettes (see \ref{vignettes} below). Table ~\ref{table:distribution_phases} summarizes participant demographics across phases.

\begin{table}[h!]
    \centering
    \begin{tabular}{l|l|l}
        \hline
        & \textbf{Phase 1} & \textbf{Phase 2} \\
        \hline
        \textbf{Education} & 50 Lower Education, 50 Higher Education & 47 Lower Education, 53 Higher Education \\
        \textbf{Gender} & 32 Female, 68 Male & 28 Female, 72 Male \\
        \textbf{Parent} & 33 Parents, 67 Not a Parent & 37 Parents, 63 Not a Parent \\
        \hline
    \end{tabular}
    \caption{Distribution of Education, Gender, and Parent Status across phases.}
    \label{table:distribution_phases}
\end{table}

\subsection{Vignette Design} \label{vignettes} 
We carefully designed vignettes to help examine how participants interpret and evaluate chatbot advice under conditions of norm divergence. Three common conditions, adult common cold, pediatric viral diarrhea, and tension headache, were selected to feature in the vignettes based on their prevalence in India and well-documented gaps in their treatment between relevant guidelines and local norms \cite{tandon2015global}. For each disease, three researchers and two local, practicing physicians created a hypothetical patient profile exhibiting a given set of symptoms. They generated two treatment plans: one that was strictly based on guidelines, and another that they knew to be guideline-divergent and norm-congruent. See Table \ref{tab:patient-profiles} for summary of our vignettes. 

\begin{table}[t]
  \caption{Symptoms and treatment vignettes associated with the 3 patient profiles.}
  \label{tab:patient-profiles}
  \centering
  \small
  \begin{tabular}{p{1.5cm}p{3.75cm}p{3.5cm}p{3.5cm}}
    \toprule
    & \textbf{Common Cold} & \textbf{Viral Diarrhea} & \textbf{Tension Headache} \\
    \midrule
    \textbf{Patient Profile} &
    35-year-old male presents with cough, rhinorrhea, low-grade fever (T $<$ 100°F), and mild myalgia. Denies sore throat. &
    1.5-year-old presents with 6–8 episodes of large-volume, watery diarrhea over the past 24 hours. Afebrile. &
    20-year-old female presents with a 3-month history of intermittent, tension-type headaches. Denies fever, emesis, or photo/phonophobia. \\
    \midrule
    \textbf{Guideline-Aligned Treatment} &
    If fever exceeds 100°F, administer Paracetamol every 6–8 hours as needed. Recommend steam inhalation, adequate rest, increased fluid intake. &
    Oral rehydration solution along with Zinc supplementation (5 mL daily) for two weeks, or advise intake of zinc-rich foods as an alternative. &
    Oral analgesics as needed in case of severe pain. Recommend maintaining adequate nutrition and sleep hygiene. \\
    \midrule
    \textbf{Norm-Congruent Treatment} &
    Initiate antibiotic therapy once daily for 5 days. Recommend steam inhalation, adequate rest, and liberal oral fluid intake. &
    Oral rehydration solution with antidiarrheal medication 1 teaspoon orally three times daily until resolution of symptoms. &
    Paracetamol injection today. Recommend maintaining adequate nutrition and sleep hygiene. \\
    \midrule
     \textbf{Context-Aware Nudge} &
    Note that the following treatment is preferred to antibiotics. Antibiotics are used to treat bacterial infections as opposed to the common cold. &
    Note that antidiarrheal medication can stop diarrhea quickly, but it is not recommended for children below two years of age. &
    Note that this medication could also be administered via an injection, but that is only recommended in acute cases. \\
    \bottomrule
  \end{tabular}
\end{table}

\begin{figure}
    \centering
    \includegraphics[width=1\linewidth, 
    alt={Fig 1. Screenshots from Whatsapp. The left pane shows the common portion of the conversation between all bots and a patient named Rahul. The right pane is split into 3 parts, showing the different treatments conferred by each bot (in summary, Verity = no medication until temp above 100*, rest, Max = antibiotics, and Clarity = "Note that the following treatment is preferred to antibiotics. Antibiotics are used to treat bacterial infections instead of. viral infections such as the common cold" no medication until temp above 100*, rest)  Each participant was shown two full side-by-side screenshots, this figure is a compressed version for reference.}]{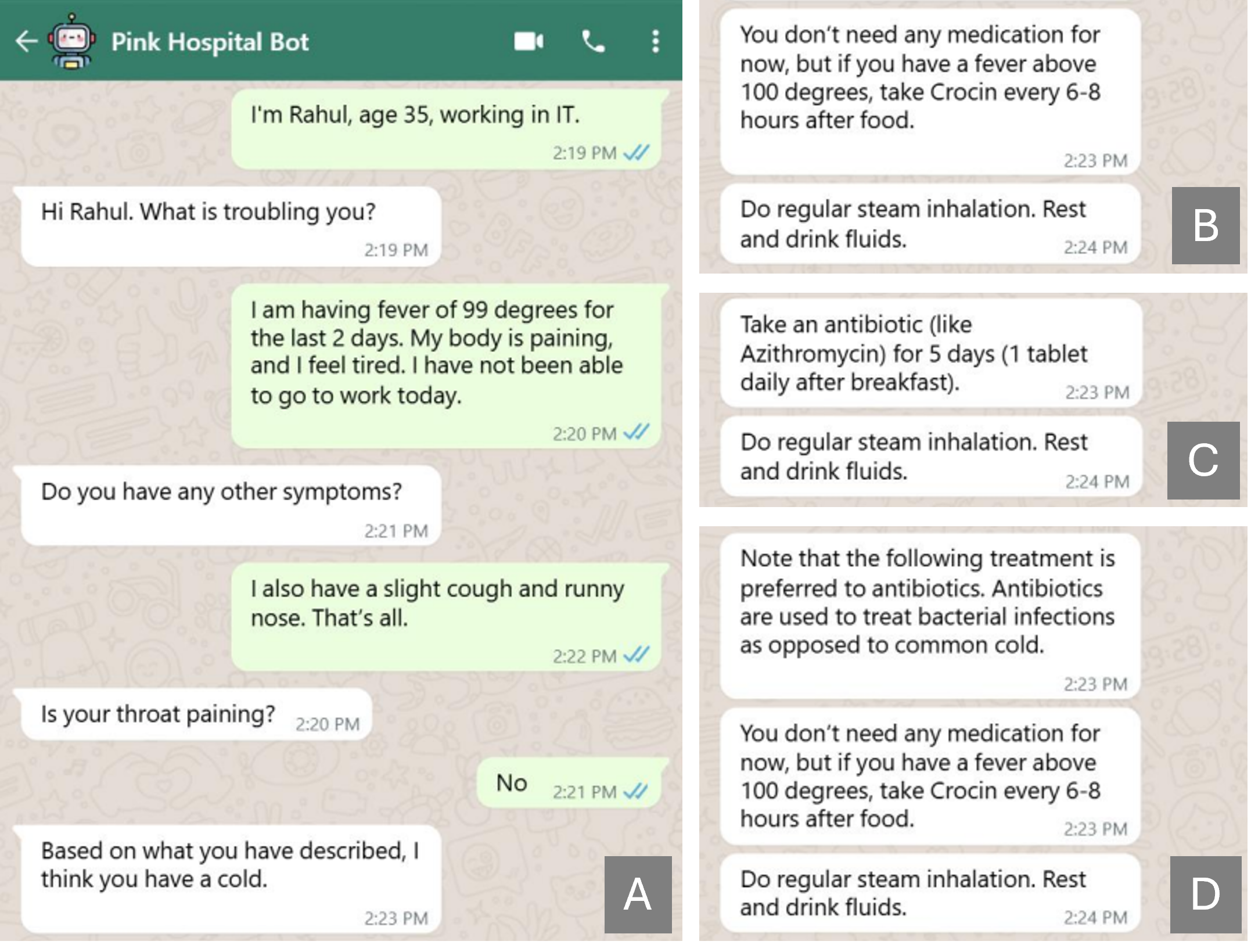}
    \caption{Pane A depicts a common message that was shown to participants across bots for the Cold scenario. On the right, pane B shows Verity's treatment, pane C shows Max's treatment message, and pane D shows Clarity's treatment message. Each participant was shown two full side-by-side screenshots, this figure is compressed for reader reference.} 
    \label{fig:bots}
\end{figure}

\newpage
Rather than deploying live chatbots to study participants, we designed standardized chatbot transcripts as artifacts based on the vignettes, enabling participants to compare and reason without the variability introduced by real-time interaction. To create the transcripts, the research team simulated conversations with three hypothetical chatbots, namely Verity, Max and Clarity. The chatbots and their implied design conditions were all anonymous when presented to participants; the names are only used to simplify the narrative in this paper. We created nearly identical conversations from the vignettes in which the patient profiles described the exact same symptoms to each of the chatbots. All three chatbots came to the same medical diagnosis after asking similar clarifying questions. The only difference in the conversations was that Verity concluded by prescribing treatment that was based on clinical guidelines, Max ended the conversation by prescribing treatment consistent with local norms, and Clarity ended the conversation with the same treatment as Verity along with a context aware nudge (see bottom row in Table \ref{tab:patient-profiles}). The context-aware nudges followed our interest in recognizing a space for the participants to dynamically construct and reconstruct meanings and legitimacy. They were iteratively designed in collaboration with the physicians to 1) foreground and acknowledge local practices, 2) proactively contextualize clinically valid advice and 3) be concise and easy to understand. The conversations between patients and the bots were simulated on WhatsApp and these screenshots of these conversations form the transcripts viewed by participants. See Figure \ref{fig:bots} for sample.

\begin{table}[h!]
    \centering
    \begin{tabular}{c|l}
        \hline
        \textbf{Bot Name} & \textbf{Treatment Style} \\
        \hline
        Verity & Presents verified medical guidance \\
        Max & Over prescribes drugs or interventions \\
        Clarity & Presents verified medical guidance with a context-aware nudge \\
        \hline
    \end{tabular}
    \caption{Bot Names and Their Treatment Styles}
    \label{table:bot_treatment_styles}
\end{table}

\subsection{Study Procedure}\label{Experiment}
Phase 1 (comparing Verity and Max) was designed to understand how participants interpreted, evaluated, and negotiated advice for chatbots that provide guideline-aligned advice versus those promoting norm-congruent advice , while Phase 2 (between Clarity and Max) was designed to understand how context-aware nudges impact users' sensemaking. Across both phases, once consent was obtained from the participant, the first author explained the study procedures and collected demographic information. Then, each participant was shown transcripts with Verity and Max, or Max and Clarity (determined based on Phase, see above) for each of the 3 diseases. Balanced variants of the experimental apparatus controlled for (1) potential confounding effects of presentation order of the bots, (2) minor conversational differences between Max and either verified bot, and (3) order of presenting the three diseases. No statistically significant differences were seen in preferences across these variations, hence we do not report stratified results. If the participant was comfortable reading by themselves, they were asked to read each pair of conversations independently. If they were not comfortable reading, we would read the conversations to them in their preferred language. Once the transcripts for each disease were known to the participant, they were asked to state Q1 (Preferred Bot): ``Which bot’s advice would you accept?,'' Q2 (Rationale): ``Please provide a justification for your selection'' and Q3 (Exposure): ``Have you or a close relative experienced <disease> in the past 6 months?'' Participants were encouraged to ask clarifying questions at any point, and the researcher remained present to support comprehension without influencing responses. The entire study procedure took under 10 minutes per participant. They were not compensated for their time due to institutional policy and local norms. 

\subsection{Analysis} 
Our analysis combined quantitative comparison of chatbot preferences with qualitative analysis of participants’ stated rationales, allowing us to examine not only what advice they gravitated towards, but how they justified their preferences. In Phase 1, we counted the number of times that the participant chose Verity across the three diseases. This value, Total, is continuous, with a minimum of 0, and a maximum of 3. Total was mapped to the participant's overall Preferred Bot in the following manner. If Verity was chosen 0 or 1 times by the participant, the participant's Preferred Bot was designated as Max. If Verity was chosen 2 or 3 times, their preferred bot was Verity. In Phase 2, we performed analogous coding based on the number of times that the participant chose Clarity across the three diseases. Preferred Bot was our binary dependent variable in each Phase. Independent variables were Educational Attainment (binary, defined as "Lower Education" representing attainment through grade twelve, and "Higher Education" representing participants with Bachelors, Masters, or PhD degrees), Gender (binary, Male or Female), Exposure (binary, Yes or No), and Parent Status (binary, Yes or No).

We chose to perform statistical analysis to accurately characterize our participants’ preferences in this phase, and allow comparison with the next phase. For both phases, we followed the same statistical analysis plan. To test whether participants chose Max more often than Verity or Clarity, we used the one sample t-test to compare the mean total to random chance. To understand the effects of education, gender, and other demographic subgroups, we used independent samples t-test and measured effect size with Cohen's d. For sub-analyses with individual disease scenarios, we again used the t-test to understand the differences between the selection of Max and Verity/Clarity. To understand the impact of context-aware nudges on bot preference across phases, we used the independent samples t-test. The two-way ANOVA test was used to understand the main and interaction effects of each pair of independent, between-subjects categorical variables (one selected from the demographics and the other being Preferred Bot). 

We used Braun and Clarke’s thematic analysis method \cite{Braun2019ReflexiveThematicAnalysis} for qualitatively coding the rationale responses from the participants. Once iterative coding was complete, themes were drawn across individual codes, and salient quotes were identified to be included in the results. Participants’ quotes, where included, are anonymized and attributed to their unique identifier (e.g., P67). 

\subsection{Ethical Considerations \& Threats to Validity}\label{Validity}
All study procedures were carried out after obtaining Institutional Review Board clearance from our institution, and written consent was obtained from individual participants. Using screenshots of the chat, as opposed to care seekers talking to a chatbot in real time, could be seen as a threat to ecological validity. However, this was a deliberate choice to ensure fair comparisons across participants by presenting standardized conversations. Also, this design avoided exposing participants currently making similar healthcare decisions to overprescriptions. We recognize that presenting overprescribing advice, even in vignette form, carries ethical risk; we therefore debriefed participants after the study, clarified which treatments aligned with clinical guidelines, and spent time answering follow up questions when they arose.

%% file: sections/4-results.tex
\section{Results}
In our study, we examine how participants interpreted and evaluated chat transcripts containing medical advice that differed in clinical recommendations, and in how those recommendations were framed. Rather than treating preference as a proxy for correctness or usability, we look at preference as an outcome of sensemaking.

Participants had similar demographics across both phases, enabling comparison of  patterns across conditions. In Phase 1, participants (N=100) had a mean age of 30.2 years (SD=9.6, range 20–58); 68\% identified as male, 33\% were parents, and educational attainment was evenly split between Lower and Higher Education groups. Phase 2 included a distinct sample of 100 participants with comparable characteristics (mean age 30.0 years, SD=7.7; 72\% male; 37\% parents; 47\% Lower and 53\% Higher Education). No participant took part in both phases.

\subsection{Phase 1: Verity vs Max}
\subsubsection{Quantitative Results}
When Verity and Max were presented to participants, a statistically insignificant majority of participants (54\%) preferred Max, i.e., they selected it on at least 2 of the 3 occasions. Twenty three percent of participants strongly preferred Max and 11\% strongly preferred Verity, selecting the given bot on all three occasions. Educational attainment shaped how participants interpreted the transcripts and their advice. Higher Education participants were significantly likely to prefer Verity, while Lower Education participants were significantly likely to prefer Max (P < .001, Cohen’s d = 0.78). Where 0 represents Max and 1 represents Verity, the Higher Education subgroup had a mean preference of M = 0.64 (SD = 0.48) while the Lower Education subgroup' mean preference was M = 0.28 (SD = 0.45). Importantly, both groups’ preferences differed significantly from random choice (Higher: t(49) = 2.04, P = .047, Cohen's d = 0.41; Lower: t(49) = 3.43, P = .001, Cohen's d = 0.69), suggesting systematic differences across groups in what constituted legitimate and acceptable medical advice. Other demographic variables (gender, parental status, prior exposure) did not significantly influence preferences.

\subsubsection{Qualitative Results}
To understand how participants arrived at these preferences, we examined their rationales. Expectations relating to 1) tangible outcomes of care, 2) impacts of given treatments and 3) behavior of a medical authority were described when constructing legitimacy. 

\paragraph{Tangible Outcomes as a Basis for Legitimacy} 
Participant sensemaking reflected a shared expectation that seeking medical advice should culminate in an intervention. Several participants explicitly compared the chatbot interaction to their experiences with in-person doctor visits, arguing that without a prescription the encounter lacked value. As P47 noted in response to Verity’s recommendation to rest, “We can take rest by ourselves. If you say no to [prescribing] a tablet, just go home, why would I take time from work and go to the doctor?” Advice to rest, in particular, was repeatedly interpreted as inaction rather than care. As P41 explained, “Action is better than doing nothing. [Verity] is too passive.” In this framing, rest failed to register as an intervention and therefore did not satisfy expectations of legitimate medical guidance, while prescriptions functioned as tangible evidence that care was underway. This made Max’s responses feel more credible and actionable, while Verity’s refusal to prescribe medication undermined its perceived usefulness. Some participants rejected Verity outright on this basis: P36 remarked that “[Verity] starts off with ‘you don’t need medicines,’ […] seems like a scam,” while P72 concluded, “If it’s [Verity], I need to go to another doctor afterwards for some real treatment; that bot is useless.” These concerns were amplified in the pediatric vignette, where not using medication was associated with risk and urgency. P69 characterized children’s illness as “an emergency situation,” while P48 emphasized the need to “give them anything to feel better quickly,” further narrowing the space in which non-interventionist advice could be recognized as acceptable care.

\paragraph{Alignment with Perceived Impact of Treatment as a Basis for Legitimacy}
Participants frequently drew on lived care experiences when making sense of chatbot advice. In doing so, they evaluated recommendations not through biomedical reasoning, but through alignment with perceived impact from prior experiences of care. As P53 remarked, “My doctor always gives antibiotics for fever, it is required.” These accounts revealed deeply entrenched beliefs about the efficacy of guideline-divergent local norms. For example, P64 asserted, “Antibiotics are a guarantee,” and P81 characterizing them as “effective for almost anything.” Such beliefs shaped interpretation even when participants acknowledged that an illness might not clinically warrant antibiotic treatment. As P25 explained, “Even if viral, we can take antibiotics, it will help. […] Antibiotics bolster the immune system.” As an aside, only two participants in Phase 1 referenced the inappropriateness of antibiotics for viral infections as a reason for preferring Verity. Similar beliefs surfaced around injections, as P60 explained, “Injection is a permanent solution, whereas pills you have to keep taking now and then.” Across these cases, participants looked to treatments and beliefs that had previously appeared to work as the primary basis for legitimacy.

\paragraph{Cues of Medical Authority as a Basis for Legitimacy}
Participants frequently assessed whether the chatbot's conversational style aligned with their expectations of how their doctors converse. In particular, concise and directive language was treated as a marker of competence, while conversational styles that offered options or invited patient decision-making were perceived as unfamiliar and therefore less credible. As P5 explained, “[Verity] gives more options to the patient unlike a typical Indian doctor which makes it seem and therefore not really believable or serious.” In contrast, Max’s language was consistently described as doctor-like, with P98 noting, “[Max] sounds more like a doctor giving [a prescription with] a big medicine name.”

Beyond this explicit standard of doctor-likeness, participants also attributed positive but loosely defined qualities to Max, describing it as “smarter overall and more human” (P14), “more understanding and trustworthy” (P6), and “more empathetic” (P33). This was despite the absence of design differences that would substantiate these traits: Max and Verity were equivalent in design and dialogue outside of the treatment recommendation, and their dialogue structures were interchanged across transcript variants and participants. In a few cases, these perceptions were strong enough to overrode agreement with Verity’s recommendations; as P19 commented, “Zinc tablets for diarrhea are known, but still [Max] seems smarter so I will accept that.” Here, norm-congruent advice appeared to retrospectively shape perceptions of competence, independent of recommendation content.

\subsection{Phase 2: Max vs Clarity}

\subsubsection{Quantitative Results}
After the incorporation of context-aware nudges, the majority of participants (85\%) showed a significant preference for Clarity (P <.001) compared to Max. Only 2\% of participants strongly preferred Max selecting it on all 3 occasions (compared to 23\% in Phase 1), while 43\% of participants now strongly preferred Clarity (compared to 11\% preferring Verity in Phase 1). This shift occurred across education groups, although there still existed a significant effect of Education on Preferred Bot (t(98)= 5.19, P <.001, Cohen's d = 0.74). Taken together, these results suggest that the nudge changed participants' sensemaking. 

\subsubsection{Qualitative Results}
In Phase 2, the context-aware nudge altered how legitimacy was constructed and contested in both participants that selected Max and Clarity. We organize qualitative findings into three themes.

\paragraph{Engaging in Reasoning as a Basis for Legitimacy} Many participants treated the context-aware nudge as a resource for engaging in or supporting reasoning about each chatbot’s legitimacy. For some, the mere presence of the nudge added credibility. As P37 noted, “Rationale is useful because I don’t know much and it shows [Clarity] is educated.” Similarly, P16 described how the explanation made the system’s intent legible, stating that Clarity was “upfront that injection is not a good idea,” which gave them “a sense for what they are thinking about when coming to a decision.” Other participants oriented to the nudge as a basis for active reasoning and scrutiny. These participants framed nudges as enabling evaluation rather than acceptance. P68 remarked, “I can know if [the treatment] is wrong through thinking about the explanation but not if there is no reason [given] at all,” while P57 echoed that Clarity’s responses “can be cross-checked.” In related accounts, participants treated the nudge as guidance that could support future sensemaking without necessarily altering their current beliefs. As P49 stated, “[Clarity] is giving a good reason, we need people to sit with us and tell us right and wrong, we do not know.”

\paragraph{Cues of Medical Authority as a Basis for Legitimacy}
 In contrast, some participants evaluated legitimacy through safely \textit{not} needing to reason. As in Phase 1, these participants preferred Max’s brevity and decisiveness, which they associated with lived experiences of care. P53 remarked that Max “sounds more like my doctor,” while P79 described Max as “short and crisp,” noting that doctors in India “have no time to tell about alternative options.” For these participants, legitimacy was tied to being able to trust an authority without needed context. As P77 reflected, “How can we judge a doctor's advice? I would take an injection if the doctor says.” In contrast to Phase 1, where Verity’s norm-divergent, minimal, advice was often met with hesitation, Max’s similarly directive advice functioned as a signal of competence rather than a limitation. Interestingly, some participants explicitly referenced the chatbot’s technological nature when articulating their sensemaking in this regard. P77 questioned, “Why is [Clarity] giving extra info… If it is a technology, it will not say wrong things. Just take the medications if [Max] is saying.” 

\paragraph{Alignment with Beliefs and Lived Constraints as a Basis for Legitimacy}
 As in Phase 1, some participants rejected Clarity’s recommendations when they conflicted with personal experience. P17 stated, “Antibiotics never have side effects. Will not trust [Clarity] because of personal experience with this condition.” Others went further into their lived experience to emphasize practical pressures that made Clarity’s guidance difficult to accept. P26 linked medication use to work demands, explaining, “Medication is needed because he has said no office for two days… there are other pressures,” point out how lack of acknowledgment of individual constraints rendered the broader context aware nudge unhelpful. P42 similarly highlighted bodily discomfort, stating, “It is hard to think about different waiting and all facts when the body is in pain.” In these cases, legitimacy was evaluated through consistency with lived realities beyond careseeking.

\paragraph{Comparative Quantitative Analysis across Phases} In Phase 1, participants did not show a statistically-significant preference between Verity and Max. In contrast, participants in Phase 2 significantly favored Clarity over Max (t(198) = 6.33, P < .001, Cohen’s d = 0.90). This shift in preference was consistent across subgroups. No interaction was found between Phase and prior Exposure to the disease (F(1,196) = 0.76, P = .38), indicating that the effect of Phase on bot preference was not moderated by prior exposure. 


%% file: sections/5-discussion.tex
\section{Discussion}
Prior work on health chatbots reports largely positive user perceptions of chatbot-delivered medical advice, including in sensitive domains \cite{Miles2021HealthChatbots, Koulouri2021ChatbotsMentalHealth}. Our findings add nuance to these conclusions by examining a setting where clinical guidelines systematically diverge from local treatment norms. Recent work in the Global South has documented widespread antibiotic self-medication and strong beliefs in the superior efficacy of injections \cite{Tripathy2018DrugPrescription, Alhassan2018InjectionPreference}. Our work extends these findings to show how such beliefs are enacted in interactions with medical chatbots. Across two phases, we demonstrate that legitimacy in our context is not contingent on clinical validity; rather, it is produced through expectation-conditioned sensemaking processes.

In Phase 1 of our work, preferences were divided between Verity, which offered guideline-aligned advice, and Max, which offered norm-congruent advice, with a slim majority (54\%) preferring Max. Importantly, these preferences were not random. Participants with lower educational attainment were more likely to prefer Max, while those with higher education more often preferred Verity, with a large effect size (Cohen’s d = 0.78). This pattern aligns with findings from health literacy literature showing that individuals with lower educational attainment often face structural gaps in access to actionable medical information \cite{PaascheOrlow2007HealthLiteracy}, while even modest educational attainment has been associated with reductions in high-risk health practices \cite{Thapa2000HighRiskChildbirth}.

Rather than portraying educational attainment as a deficit, our qualitative findings in Phase 1 suggest that participants drew on different sensemaking strategies when evaluating advice, with three expectations needing to align in order to signal legitimacy: (1) the production of a tangible outcome, (2) consistency with perceived of treatment, and (3) resemblance to familiar conversation styles of Indian doctors. In our sample, recommendations that did not produce a tangible outcome were treated as incomplete or unsatisfactory, regardless of their clinical validity. Worryingly, prior work suggests that clinicians similarly orient to perceived patient expectations, contributing to higher rates of guideline-divergent prescribing \cite{Wagner2024WhatDrivesPoorQuality}. Importantly, our participant expectations persisted even in an online setting where participants did not incur the usual costs of time, travel, or consultation. This suggests that expectations for tangible outcomes are anchored less in the economics of care-seeking than in norms governing what legitimate medical encounters should produce. At the same time, the increasing availability of low- or no-cost medical chatbots may begin to decouple advice from immediate financial exchange, creating space for norms to be renegotiated over time. Whether chatbots can gradually shift prescription expectations toward clinically validated norms, and whether this, in turn, might reduce pressure on physicians to overprescribe, remains an open and important area for future work. Such shifts would not reflect changes in knowledge alone, but changes in the cues through which legitimacy is established.

In terms of alignment with prior beliefs, only two participants out of one hundred cited the rationale that antibiotics should not be taken for viral conditions. This asymmetry suggests that clinical understanding does not factor heavily in enacting legitimacy, especially when it conflicts with treatments participants have experienced as effective in the past. This marks a departure from Western contexts where signaling guideline-alignment positively impacts perceived legitimacy, and affirms work finding that social norms shape what feels trustworthy \cite{sharifa, Wadhwa2025DesigningWithCulture}. Beyond treatment content, participants consistently projected expectations of human doctors onto chatbots. They inferred authority from familiar interactional styles, and went as far as to construct differences in perceived competence between the two chatbots despite their behavioral equivalence. Taken together, these findings indicate that users actively interpret, negotiate, and at times resist chatbot advice based on cues that are specific to their local context.

In Phase 2, we introduced context-aware nudges to understand how they could create space for or reshape sensemaking. A significant majority of participants (85\%) favored Clarity. Their rationale underscored the nudge as an interactional resource, allowing participants engage in understanding, reflection, and learning, rather than inferring legitimacy solely through expectation alignment. When Verity had previously deviated from these expectations without explanation, it risked being read as passive, unintelligent, or illegitimate. When Clarity supported users in understanding why a different course of action may be appropriate, participants were more willing to accept unfamiliar treatments. That this pattern held across gender, parental status, and prior exposure suggests that interactional framing (rather than demographic targeting) plays a central role in reconciling norm-divergent advice with lived experience. This result is promising, particularly given that the nudges we designed were far milder than fear-based messaging previously used to shift health behaviors in other contexts \cite{Sirota2024PostAntibioticFuture}.

At the same time, explanation was not universally welcomed. Some participants preferred Max precisely because it resembled brief, transactional encounters with doctors, framing authority as something to be deferred to rather than interrogated. Others experienced the additional information as cognitively overwhelming, mirroring prior work \cite{foo}. These findings surface a tension in the design of medical chatbots: nudges or scaffolding can support sensemaking, but they can also impose uneven cognitive burdens, particularly in contexts shaped by inequality. Although preferences in Phase 2 shifted decisively toward Clarity across education groups, a moderate education effect persisted (Cohen’s d = 0.74), suggesting that explanation alone may not fully resolve equity challenges if participants with lower educational attainment continue to gravitate toward familiar but guideline-divergent care. 

Our findings indicate that sensemaking is contingent on how context is invoked and made available to users. But building nudges requires explicitly defining a “context” for a given user, and perhaps interrogating whether any answer can exist in the absence of (an implicit) context. Beyond socioeconomic status and location, there are a whole host of personal biases and beliefs that may be relevant for chatbots to consider if they want to be adopted. For example, personal preferences around delaying medication use, or choosing to intervene immediately may be relevant context that impacts perceived legitimacy of a given treatment. At the same time, expanding context awareness introduces familiar tensions: highly contextualized responses may invite scrutiny, raise ethical and privacy concerns, or be experienced as intrusive \cite{Svikhnushina2021UserExpectations}. From a DIS perspective, this underscores the need for designs that balance context-aware interpretation support with privacy-preservation and restraint.

Taken together, our findings position medical chatbots as active participants in the negotiation of authority within healthcare systems. When chatbots mirror norms, they risk stabilizing potentially harmful norms that can have population-level consequences. When they are used to expand access to care, there may be an interest in bringing patients towards safer practices.  When chatbots challenge local practices, they risk introducing uncertainty and may shift responsibility for clinical judgment onto users lacking adequate support in interpretation \cite{sharifa}. Moreover, when any systems implicitly position themselves as normative arbiters of what is “accurate” care, they may erode trust in local providers or delegitimize experiential knowledge that patients rely on to navigate constrained health systems. Designing for critical engagement raises broad design questions about medical chatbots in the Global South: Can \textit{and should} technology intervene in the cycle of expectation and legitimacy in health care provision? If so, how can interaction design safely support user reinterpretation of legitimate care? We are hopeful that by foregrounding these questions, we pave the way for DIS researchers to critically analyze the next generation of healthcare technologies that are meant to support and care for marginalized people. Only with a holistic, end-to-end view of the eventual beliefs, behaviors and outcomes precipitated by chatbots will we be able to appreciate the full scope of their potential benefits, limitations and harms.

\section{Limitations} Although we made attempts to reduce the skew in our participant cohorts, we used convenience sampling in our recruitment, which may introduce selection bias and affect the representativeness of the sample. Further, we worked with a population in urban India, and views of chatbots and their advice may be different in rural/periurban contexts and diverse cultures.

%% file: sections/6-conclusion.tex
\section{Conclusion}
Our two-phase, mixed-methods analysis shows that guideline alignment alone is not a sufficient signal for medical chatbots to be perceived as legitimate in contexts where lived treatment norms may diverge from the guidelines. Through a vignette-based study conducted in two phases in India, we demonstrate that users’ interpretations of medical advice are shaped by expectations of intervention, cues of medical authority, and consistency with prior care experiences. When guideline-aligned advice is presented without contextual framing, many participants reject it in favor of norm-congruent recommendations. By contrast, incorporating context-aware nudges shifts preferences toward guideline-aligned advice by supporting users’ sensemaking. Taken together, these findings surface tensions that complicate the development of medical AI systems and point to the need for design strategies that can reconcile chatbot-delivered care with diverse local norms in Global South settings.